\documentclass[12pt]{paper}
\usepackage{geometry}
\usepackage{amsfonts}
\usepackage{amsmath}
\usepackage{amssymb}
\usepackage{graphicx}
\usepackage{mathrsfs} 
\usepackage{color} 

\providecommand{\tanh}{\mathrm{th}}

\newcommand{\pd}[2]{\frac{\partial #1}{\partial #2}}

\title{Waves in slowly varying band-gap media} 
\author{
Ory Schnitzer\thanks{o.schnitzer@imperial.ac.uk}\\[0.25cm] \normalsize{Department of Mathematics, Imperial College London, London SW7 2AZ, UK}}

\date{}
\usepackage{amsopn}

\begin{document} 

\maketitle

\begin{abstract}
This paper is concerned with the asymptotic description of high-frequency waves in locally periodic media. A key issue is that the Bloch-dispersion curves vary with the local microstructure, giving rise to hidden singularities associated with band-gap edges and branch crossings. We describe an asymptotic approach for overcoming this difficulty, and take a first step by studying in detail the simplest case of 1D Helmholtz waves. The method entails matching adiabatically propagating Bloch waves, captured by a multiple-scale Wentzel--Kramers--Brillouin (WKB) approximation, with complementary multiple-scale solutions spatially localised about dispersion singularities. Within the latter regions the Bloch wavenumber is nearly critical; this allows their homogenisation, following the method of high-frequency homogenisation (HFH), over a naturally arising scale intermediate between the periodicity (wavelength) and the macro-scale. Analogously to a classical turning-point analysis, we show that close to a spatial band-gap edge the solution is an Airy function, only that it is modulated on the short scale by a standing-wave Bloch eigenfunction. By carrying out the asymptotic matching between the WKB and HFH solutions we provide a detailed description of Bloch-wave reflection from a band gap. Finally, we implement the asymptotic theory for a layered medium and demonstrate excellent agreement with numerical computations. 
\end{abstract}

%

\section{Introduction}
Waves in periodic media is a vast subject of great interest in numerous fields. Classical methods of averaging and mathematical homogenisation are useful in applications where the wavelength is large relative to the periodicity. There are many fields, however, where the r\'egime of interest concerns wavelengths commensurate with the periodicity, as in gratings, layered media, photonic (similarly phononic, platonic,...) crystals and solid-state quantum mechanics \cite{Sakoda:Book,Joannopoulos:Book}. In these applications, the basic theoretical constituents are Bloch-wave solutions, i.e., periodically modulated plane waves supported by an infinitely periodic medium \cite{Brillouin:Book}. Dispersion surfaces relating the Bloch eigenfrequencies and eigen-wavevectors encapsulate a breadth of information including the rate and direction in which energy may propagate \cite{Notomi:10} and the existence of frequency band gaps wherein propagation is forbidden \cite{Yablonovitch:87}.

An unbounded, exactly periodic, medium is an idealisation and applications usually involve crystal boundaries, tapered designs or defects. Fortunately, there is often a scale separation between the periodicity (wavelength) and a ``macro-scale'' associated with crystalline boundaries, the length on which the crystal is tapered, the width of an incident beam, or the position of a source. It is then intuitive that the wave field is locally, in some approximate sense and away from boundaries, sources and singularities, a finite superposition of Bloch solutions. It is possible to intuitively build on this picture by encoding the local Bloch-dispersion characteristics in a Hamiltonian model \cite{Russel:99,Cassan:11}. 
The equations of motion determine the variation of the Bloch wavevector along particle paths, or ``Bloch rays'', which trace the group-velocity vector. The group velocity linearly relates the macroscopic energy flux and density \cite{Sakoda:Book}; hence, given the local ray pattern and Bloch eigenmodes, the change in amplitude along a ray follows from energy conservation. Furthermore, analysing the iso-frequency contours of the dispersion surfaces in conjunction with Bragg's law allow identifying reflected, refracted (and high-order diffracted) Bloch rays at crystalline interfaces \cite{Zengerle:87,Joannopoulos:Book} (their amplitudes and phases, however, sensitively depend on details of the interface). This physical picture has been invaluable in studies of transmission through finite photonic- \cite{Notomi:10,luo:02} and platonic- \cite{Smith:12} crystal slabs, and slowly varying photonic and sonic crystals designed for cloaking and steering \cite{Kurt:08,Urzhumov:10,Romero:13}. It is difficult, however, to systematically cast these ideas into a detailed quantitative approximation of the wave field, especially when crystalline interfaces and dispersion singularities are important. Recent studies have also highlighted the role played by the Berry phase \cite{Berry:84} in graded \cite{Johnson:02} and symmetry-breaking crystals \cite{Raghu:08}, and slowly varying waveguides \cite{Tromp:92}; the Berry phase is sometimes incorporated in Hamiltonian formulations through a small ``anomalous velocity'' \cite{Panati:03,Raghu:08,Weinan:13}.

It is well known that traditional ray theory (i.e., for single-scale materials) can be derived as the short-wavelength asymptotic limit of the wave equation \cite{Holmes:Book}. Analogously, a suitable starting point for an asymptotic theory of Bloch rays is a \emph{multiple-scale} Wentzel--Kramers--Brillouin (WKB) ansatz \cite{Bensoussan:11}. It may seem surprising, given the huge applicability of geometric optics, that works along these lines have largely focused on rigorous proofs in the context of propagation and localisation of wave packets \cite{Allaire:10,Allaire:11,Cherednichenko:15,Harutyunyan:16}. These analyses are compatible with the intuitive notion of Bloch rays; they show, however, that slow Berry-like phases are important even when seeking only a leading-order approximation to the wave field. These analyses disregard diffraction at crystalline boundaries and caustics, and the dominant Bloch eigenfrequencies are assumed to be within a pass band, away from singularities in the dispersion landscape.  It seems that, to date, there has been little effort to implement and pragmatically extend these ideas towards a more applicable approximation scheme. 

Dispersion singularities, namely critical points at the edges of complete and partial band gaps, along with branch crossings, are responsible for phenomena such as slow light, dynamic anisotropy \cite{Notomi:00,Ceresoli:15}, unidirectional propagation \cite{Colquitt:16} and non-trivial topological invariants \cite{Lu:14}. Fortunately, wave propagation at near-critical frequencies is well described by a distinct multiple-scale asymptotic scheme known as High-Frequency Homogenisation (HFH) \cite{Craster:10}. HFH furnishes a (typically unimodal) homogenised description where the wave field is provided as a product of an envelope function, governed by a macro-scale equation, and a rapidly varying Bloch eigenfunction, governed by the Bloch cell problem at the nominal (typically standing-wave) frequency. The method has been extensively applied to study photonic \cite{Antonakakis:13,Maling:15} (similarly phononic \cite{Antonakakis:14,Colquitt:15high}, platonic \cite{Antonakakis:12}) crystals, Rayleigh--Bloch waves along gratings \cite{Colquitt:15rayleigh}, and layered media \cite{Joseph:15}. The method is a multiple-scale analogue of near-cut-off expansions in waveguide theory \cite{Craster:09} and is also intimately connected to the KP perturbation method and the idea of an effective mass in solid-state physics \cite{Kittel:Book,Naraigh:12}. 

The frequency intervals covered by multiple-scale WKB expansions and HFH complement each other. In fact, the respective regions of validity overlap at intermediate perturbations away from critical frequencies. Naively, it may appear that only one of these two methods is necessary at any given frequency. This is certainly incorrect for a graded crystal where the dispersion surfaces vary on a macro-scale and may easily become singular. Indeed, Johnson \textit{et al.}~warn that this is a ``pitfall to avoid in the design of tapered photonic crystal waveguides'' \cite{Johnson:02}; an otherwise adiabatically propagating Bloch wave would strongly reflect from the singularity. While less evident, monochromatic dispersion singularities are also ubiquitous in uniform crystals, at least in two and three dimensions. For instance, if the operating frequency lies in a partial band gap, propagation is forbidden in certain directions \cite{Ceresoli:15}; there are then singular Bloch rays separating bright and shadow regions along which the frequency is critical. To date, asymptotic analyses have exclusively focused on either non-critical regions, using the WKB method or related approaches \cite{Johnson:02}, or studied scenarios where the frequency is everywhere nearly critical, using HFH. We argue here that appropriately combining the two methods provides an avenue to a more complete asymptotic theory of high-frequency waves in locally periodic media, covering all dispersion branches, both near and away from  dispersion singularities. Specifically, we anticipate that WKB Bloch waves can be asymptotically matched in space with localised HFH-type expansions describing intermediate \emph{length} scales about dispersion singularities. 

In this paper we demonstrate this approach in the simplest case of time-harmonic Helmholtz waves in a 1D locally periodic medium. We begin in \S\ref{sec:bulk} by considering the ``outer'' problem where Bloch waves propagate in a pass-band region, adiabatically varying together with the local dispersion relation of the medium. Our multiple-scale WKB analysis differs from previous works in several aspects. First, we work in the frequency domain. Second, we decompose the complex-valued transport equation into two simpler real equations respectively governing the amplitude and a slow phase; moreover, by considering perturbations of the Bloch eigenvalue problem we derive exact identities that allow us to simplify the asymptotic formulae. Third, we consider a more general multiple-scale representation of the locally periodic medium, which is consistent also with layered media.
In \S\ref{sec:behaviour} we investigate the divergence of the WKB description as a spatial band-gap edge singularity is approached. In \S\ref{sec:hfh} we study the region close to a band-gap edge using a HFH-like expansion. In \S\ref{sec:reflection} we asymptotically match the WKB and HFH expansions in the scenario where a Bloch wave is reflected from a band-gap edge. In \S\ref{sec:layered} we demonstrate the validity of our approach by comparing the asymptotic results with numerical computations in the case of a layered medium. In the concluding section \S\ref{sec:discussion} we anticipate generalisations and extensions to this work.

\section{Slowly varying Bloch waves}\label{sec:bulk}
Consider the reduced wave equation
\begin{equation} \label{master eq}
\epsilon^2\frac{d^2 u}{dx^2}+{\omega^2}\Lambda u=0.
\end{equation}
In Eq.~\eqref{master eq}, $x$ is a spatial coordinate normalised by a long scale $L$, and $\omega$ denotes the frequency normalised by $c/l$, where $c$ is a reference wave speed and $l$ is a short length scale such that $\epsilon=l/L\ll1$. Our interest is in a locally periodic medium described by the real function $\Lambda=\Lambda[\xi,r(x)]$, which is an arbitrary $2$-periodic function of the short-scale coordinate $\xi=x/\epsilon$ and a smooth function of the long scale $x$ through the parameter $r$. Specifically, our interest is in the limit where $\epsilon\to0$, with $\Lambda,\omega=O(1)$. This limit process corresponds to a high-frequency r\'egime where wavelength and periodicity are commensurate and small compared to the long scale $L$; we emphasise that ${|\Lambda-1|}$ is not assumed to be small. 

As explained in \S\ref{sec:layered}, in the case of a layered medium it is necessary to assume $\Lambda$ has an expansion in powers of $\epsilon$. For the sake of calculating the wave field $u$ to leading order, however, only the first \emph{two} terms in this expansion are relevant:
\begin{equation}\label{Lambda def}
\Lambda = \chi[\xi,r(x)]+\epsilon \tau[\xi,r(x)].
\end{equation}
Here $\chi,\tau$ are real, $\epsilon$-independent, $2$-periodic functions of $\xi$, and smooth functions of $x$ through the parameter $r(x)$. 

\begin{figure}
\begin{center}
\includegraphics[scale=0.33]{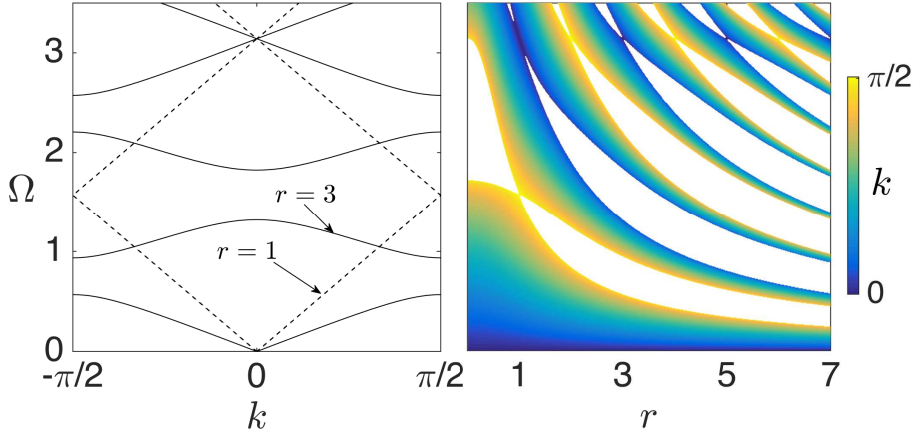}
\caption{Bloch dispersion relation for the layered medium considered in \S\ref{sec:layered}.}
\label{fig:disp}
\end{center}
\end{figure}

In this section we seek asymptotic solutions in the form of a multiple-scale WKB ansatz
\begin{equation} \label{master}
u = A(\xi,x)e^{i\varphi(x)/\epsilon},
\end{equation}
where $\varphi$ is real and $A$ is complex and $2$-periodic in $\xi$. Substituting \eqref{master} into \eqref{master eq} yields
\begin{multline}\label{Aeq}
\pd{^2A}{\xi^2}+2i \frac{d\varphi}{dx}\pd{A}{\xi}+\left[\omega^2\Lambda-\left(\frac{d\varphi}{dx}\right)^2\right]A \\
+ \epsilon\left(2\pd{^2A}{x\partial \xi}+2i\frac{d\varphi}{dx}\pd{A}{x}+i\frac{d^2\varphi}{dx^2}A\right)+\epsilon^2\pd{^2A}{x^2}=0.
\end{multline}
The form of \eqref{Aeq} suggests expanding $A$ as
\begin{equation} \label{A exp}
A(\xi,x) \sim A_0(\xi,x)+\epsilon A_1(\xi,x) + \cdots, 
\end{equation}
where $A_0,A_1,\ldots=O(1)$ are complex-valued and $2$-periodic in $\xi$. The $O(1)$ balance of \eqref{Aeq} is:
\begin{equation}\label{A0eq}
\mathscr{L}^2_{k,\chi}A_0=0,
\end{equation}
where we define the $x$-dependent operator $\mathscr{L}^2_{k,\chi}$ and wavenumber $k$ as
\begin{equation}\label{L2}
\mathscr{L}^2_{k,\chi}=\pd{^2}{\xi^2}+2i k\pd{}{\xi}+\left(\omega^2\chi-k^2\right), \quad k(x) = \frac{d\varphi}{dx}.
\end{equation}
For $x=x'$ fixed, \eqref{A0eq} together with periodicity conditions constitutes the familiar Bloch eigenvalue problem \cite{Joannopoulos:Book} with periodic modulation $\chi[\xi,r(x')]$ and Bloch wavenumber $k(x')$. Thus, standard techniques can be employed in order to calculate a dispersion relation 
\begin{equation}\label{disp}
\Omega(k,r)=\omega,
\end{equation}
where $\Omega$ is multivalued. As an example, the dispersion relation for the layered medium discussed in \S\ref{sec:layered} [cf.~\eqref{layered chi}] is visualised in Fig.~\ref{fig:disp}; note the existence of frequency band gaps, wherein $k$ is complex valued, and  crossing points at higher frequencies, where $k$ is a degenerate eigenvalue. In this section, we assume $x$ lies within a pass band, namely that the eigenvalue $k$ is real and non-degenerate. Without loss of generality, we only need to consider wavenumbers in the first Brillouin zone, $-\pi/2\le k\le \pi/2$. Let $U(\xi,\{k,r\})$ denote the eigenfunctions, which satisfy $\mathscr{L}^2_{k,\chi}U=0$ and are $2$-periodic in $\xi$. From time-reversal symmetry, if $k$ is an eigenvalue then so is $-k$ and $U(\xi,\{-k,r\})={U}^*(\xi,\{k,r\})$, where an asterisk denotes complex conjugation. In the appendix we derive the result \cite{Sakoda:Book}
\begin{equation}\label{gv relation}
\pd{\Omega}{k}= \frac{k\int|U|^2\,d\xi-i\int  U^*\pd{U}{\xi}\,d\xi}{\omega\int\chi|U|^2\,d\xi},
\end{equation}
which gives the group velocity in terms of integrals of the eigenfunctions over an elementary cell, say $\xi\in[-1,1)$.

The above discussion suggests writing the solution of \eqref{A0eq} as
\begin{equation}\label{A form}
A_0 = f(x)U\left(\xi;\{k,r\}\right),
\end{equation}
where $f$ is complex valued; it will be convenient to treat the amplitude $|f|$ and phase $\gamma$ of $f$ separately, 
\begin{equation}
f(x)=|f(x)|e^{i\gamma(x)}.
\end{equation}
It is important at this stage to note that the eigenfunctions $U$ are determined for a fixed $x$ only up to an arbitrary complex-valued prefactor. Indeed, the product $fU$ remains invariant under the gauge transformation 
\begin{multline}\label{gauge fU}
U\left(\xi;\{k,r\}\right)\rightarrow a(x)e^{ib(x)}\acute{U}\left(\xi;\{k,r\}\right), \\ |f(x)| \rightarrow \frac{|\acute{f}(x)|}{a(x)}, \quad \gamma(x) \rightarrow \acute{\gamma}(x) - b(x),
\end{multline}
where $a(x)$ and $b(x)$ are real.
We shall assume that a specific set of eigenfunctions $U$, varying smoothly with $x$, has been chosen.

Consider now the $O(\epsilon)$ balance of \eqref{Aeq}:
\begin{equation}\label{A1eq}
\mathscr{L}^2_{k,\chi}A_1 + 2\pd{^2A_0}{x\partial\xi}+2ik\pd{A_0}{x}+i\frac{dk}{dx}A_0 +\omega^2\tau A_0= 0.
\end{equation}
Eq.~\eqref{A1eq} is a forced variant of \eqref{A0eq} and we derive a solvability condition by multiplying \eqref{A1eq} by ${A}^*_0$, subtracting the conjugate of \eqref{A1eq} multiplied by $A_0$, and integrating by parts over the unit cell $\xi\in[-1,1)$. Noting the periodicity of $A_0$ and $A_1$  we find the transport equation:
\begin{equation}\label{solv complex}
-2\int \pd{{A}^*_0}{\xi}\pd{A_0}{x}\,d\xi + 2ik\int \pd{A_0}{x}{A}^*_0\,d\xi
+i\frac{dk}{dx}\int|A_0|^2\,d\xi+\omega^2\int\tau|A_0|^2\,d\xi =0.
\end{equation}
Inspection of the imaginary and real parts of \eqref{solv complex} leads to more useful equations separately governing the ``amplitude'' $|f|$ and the ``slow phase'' $\gamma$, respectively. Thus, subtracting from \eqref{solv complex} its conjugate, and using \eqref{gv relation}, we find the energy balance
\begin{equation}\label{amp eq}
\frac{d}{dx}\left(\pd{\Omega}{k}|f|^2\int\chi|U|^2\,d\xi\right) = 0.
\end{equation}
The converse manipulation yields
\begin{equation}\label{gamma A}
\mathrm{Re}\left[\int\left(\pd{A_0}{\xi}+ikA_0\right)\pd{A_0^*}{x}\,d\xi\right]=\frac{\omega^2}{2}\int\tau|A_0|^2\,d\xi,
\end{equation}
which, by eliminating $|f|$ using \eqref{A form} and \eqref{gv relation}, gives an equation for $\gamma$:
\begin{equation}\label{gamma}
\frac{d\gamma}{dx}\pd{\Omega}{k}\omega\int\chi|U|^2\,d\xi = -\mathrm{Re}\int\left(\pd{U}{\xi}+ikU\right)\pd{ U^*}{x}\,d\xi+\frac{\omega^2}{2}\int\tau|U|^2\,d\xi.
\end{equation}
By substituting the two identities \eqref{dk} and \eqref{dU}, derived in the appendix, \eqref{gamma} can be rewritten in a more convenient form where differentiation with respect to $\xi$ is circumvented:
\begin{equation}\label{gamma2}
\frac{d\gamma}{dx}=\frac{\mathrm{Im}\int\pd{\chi}{r}\pd{U^*}{x}U\,d\xi}{\int\pd{\chi}{r}|U|^2\,d\xi}+\frac{\omega\int\tau|U|^2\,d\xi}{2\pd{\Omega}{k}\int\chi|U|^2\,d\xi}.
\end{equation}

The asymptotic solution in a spatial pass band may also involve an oppositely propagating Bloch wave. A general leading-order approximation is therefore
\begin{equation}\label{WKB solution}
u \sim \mathcal{A}_+\frac{Ue^{i\gamma^+}e^{i\varphi^+/\epsilon}}{\sqrt{\left|\pd{\Omega}{k}\right|\int{\chi}|U|^2\,d\xi}}
+\mathcal{A}_-\frac{ U^*e^{i\gamma^-}e^{i\varphi^-/\epsilon}}{\sqrt{\left|\pd{\Omega}{k}\right|\int{\chi}|U|^2\,d\xi}},
\end{equation}
where
\begin{equation}\label{phi pm}
\frac{d\varphi^{\pm}}{dx}=\pm k(x), 
\end{equation}
\begin{equation}\label{gamma pm}
\frac{d\gamma^{\pm}}{dx}=\pm\frac{\mathrm{Im}\int\pd{\chi}{r}\pd{U^*}{x}U\,d\xi}{\int\pd{\chi}{r}|U|^2\,d\xi}\pm\frac{\omega\int\tau|U|^2\,d\xi}{2\pd{\Omega}{k}\int\chi|U|^2\,d\xi},
\end{equation}
and  $\mathcal{A}_{\pm}$ are complex constants. 
In writing \eqref{gamma pm} we note that the group velocity is odd in $k$ and that under the transformation $k\rightarrow-k$ the first integral on the right-hand side of \eqref{gamma} is conjugated. We further note that the direction of propagation of the WKB solutions in \eqref{WKB solution} is determined by the sign of the group velocity rather than that of $k$.

\section{Breakdown of the WKB description near a band-gap edge}\label{sec:behaviour}
In this section we investigate the manner in which the WKB solutions \eqref{WKB solution} loose validity near a band-gap edge. Thus, let $x=x_t$ divide the macro-scale into spatial regions wherein $\omega$ respectively falls within a pass band ($k$ real) and a stop band ($k$ complex). We denote $r(x_t)=r_t$ and $k(x_t)=k_t$, and assume the usual case where the critical wavenumber is at the centre or edge of the Brillouin zone (see Fig.~\ref{fig:disp}), i.e., $k_t=0$ or $\pi/2$, respectively. Assuming $r(x)$ is not critical at $x_t$ we write
\begin{equation}\label{r exp}
r(x) \sim r_t + (x-x_t)\left(\frac{dr}{dx}\right)_t + O(x-x_t)^2 \quad \text{as} \quad x\to x_t
\end{equation}
and
\begin{equation}\label{chi exp}
\chi(\xi,r) \sim \chi_t(\xi) + (x-x_t)\left(\pd{\chi}{r}\right)_t\left(\frac{dr}{dx}\right)_t+ O(x-x_t)^2 \quad \text{as} \quad x\to x_t,
\end{equation}
where a $t$ subscript denotes evaluation at $x=x_t$. 
Time-reversal symmetry and periodicity of the dispersion curves suggest that
\begin{equation}\label{zero gv}
\left(\pd{\Omega}{k}\right)_t = 0, \quad \left(\pd{^2\Omega}{k\partial r}\right)_t=0, \quad \left(\pd{^3\Omega}{k^3}\right)_t=0, \quad \ldots,
\end{equation}
which in turn imply the expansion
\begin{multline}\label{Omega exp B}
\Omega(r,k) \sim \Omega_t + (r-r_t)\left(\pd{\Omega}{r}\right)_t   +  \frac{1}{2}(k-k_t)^2\left(\pd{^2\Omega}{k^2}\right)_t \\ + O[(r-r_t)^2,(r-r_t)(k-k_t)^2,(k-k_t)^4] \quad \text{as} \quad x\to x_t.
\end{multline}
But $\Omega(r,k)=\omega$ for all $x$, hence
\begin{equation}\label{k exp}
k(x)\sim k_t + k_{1/2}(x-x_t)^{1/2}+ O\left(|x-x_t|^{3/2}\right) \quad \text{as} \quad x\to x_t,
\end{equation}
where
\begin{equation}\label{k12 thermo}
k_{1/2}^2=-2{\left(\frac{dr}{dx}\right)_t\left(\pd{\Omega}{r}\right)_t}\Big{/}{\left(\pd{^2\Omega}{k^2}\right)_t}.
\end{equation}
In \eqref{k exp} and throughout the paper the square-root function has a branch cut along the negative imaginary axis and is positive along the positive real axis. In \eqref{k12 thermo}, the sign of $k_{1/2}$ is chosen so that $k-k_t$ has the correct sign when $x$ is on the side of $x_t$ where $k$ is real. It follows from \eqref{Omega exp B} and \eqref{k exp} that 
\begin{equation}\label{gv expansion}
\pd{\Omega}{k} \sim  k_{1/2}(x-x_t)^{1/2}\left(\pd{^2\Omega}{k^2}\right)_t + O(x-x_t) \quad \text{as} \quad x\to x_t.
\end{equation}

Having the above expansions for the dispersion characteristics of the material near $x_t$, we are ready to obtain the corresponding asymptotics of \eqref{WKB solution}. Consider first the fast phases $\varphi^{\pm}$; from \eqref{phi pm} together with \eqref{k exp} we find
\begin{equation}
\varphi^{\pm}(x) \sim \varphi^{\pm}_t \pm k_t (x-x_t) \pm \frac{2}{3}k_{1/2}(x-x_t)^{3/2} + O\left(|x-x_t|^{5/2}\right)  \quad \text{as} \quad x\to x_t. 
\end{equation}
Subtler are the asymptotics of $U$ and $\gamma^{\pm}$, particularly because $U$ and $\gamma$, considered separately, are ``gauge dependent''  [cf.~\eqref{gauge fU}]. Fixing the standing-wave eigenfunction $U(\xi,{k_t,r_t})=U_t(\xi)$, the perturbation theory in the appendix gives
\begin{equation}\label{U exp}
U(\xi,\{k,r\}) \sim U_t(\xi) + (x-x_t)^{1/2}U_{1/2}(\xi) + \cdots \quad \text{as} \quad x\to x_t.
\end{equation}
As explained in the appendix, $U_{1/2}$ remains gauge dependent even after fixing the standing-wave eigenfunction; in fact, in general one could add to \eqref{U exp} an intermediate-order term proportional to $U_t$. We assume the chosen gauge entails no such intermediate term, which would only arise in an analytical or computational scheme if intentionally forced to. 
Substituting \eqref{U exp} and  \eqref{gv expansion} into \eqref{gamma pm} shows that the $x$ derivatives of $\gamma^{\pm}$ are $O\left(|x-x_t|^{-1/2}\right)$ as $x\to x_t$, which implies
\begin{equation}\label{gamma exp}
\gamma^{\pm} (x) \sim \gamma^{\pm}_t + O\left(|x-x_t|^{1/2}\right) + \cdots \quad \text{as} \quad x\to x_t.
\end{equation}

The asymptotic estimates \eqref{gv expansion}, \eqref{U exp} and \eqref{gamma exp} together yield the behaviour of the \emph{leading-order} WKB solution \eqref{WKB solution}:
\begin{multline}\label{WKB asym}
u \sim \frac{\mathcal{A}_+ e^{i\gamma_t^+ + i\varphi_t^+/\epsilon}}{\sqrt{\left|k_{1/2}\left(\pd{^2\Omega}{k^2}\right)_t\right|\int{\chi_t}|U_t|^2\,d\xi}}\left\{\frac{U_t(\xi)}{|x-x_t|^{1/4}}+O\left(|x-x_t|^{1/4}\right)\right\} \\
\times \exp\left\{ ik_t \frac{x-x_t}{\epsilon} + \frac{2}{3}i k_{1/2}\frac{(x-x_t)^{3/2}}{\epsilon} + O\left(|x-x_t|^{5/2}/\epsilon\right)\right\} \\
+ \frac{\mathcal{A}_- e^{i\gamma_t^- + i\varphi_t^-/\epsilon}}{\sqrt{\left|k_{1/2}\left(\pd{^2\Omega}{k^2}\right)_t\right|\int{\chi_t}|U_t|^2\,d\xi}}\left\{\frac{\bar{U}_t(\xi)}{|x-x_t|^{1/4}}+O\left(|x-x_t|^{1/4}\right)\right\} \\
\times \exp\left\{ -ik_t \frac{x-x_t}{\epsilon} - \frac{2}{3}i k_{1/2}\frac{(x-x_t)^{3/2}}{\epsilon} + O\left(|x-x_t|^{5/2}/\epsilon\right) \right\} \quad \text{as} \quad x\to x_t.
\end{multline}
Expansion \eqref{WKB asym} can be made more revealing by noting that $U_t\exp(ik_t\xi)$ is anti-periodic and real up to a multiplicative complex prefactor. We may therefore write $U_t=\exp(i\beta)\exp(-ik_t\xi)\mathcal{U}_t(\xi)$, where $\beta$ is a real phase depending on the gauge choice and $\mathcal{U}_t(\xi)$ is a periodic (if $k_t=0$) or anti-periodic (if $k_t=\pi/2$) real function; similarly, $U^*_t=\exp(-i\beta)\exp(ik_t\xi)\mathcal{U}_t(\xi)$. We next note that $\exp(ik_t\xi)\exp(-ik_t x/\epsilon)$ and $\exp(-ik_t\xi)\exp(ik_t x/\epsilon)$ represent exactly the same function. Note that, for $k_t=\pi/2$, the latter function is \emph{not} the unity function, since the dependence on $\xi$ is extended from the unit cell $\xi\in(-1,1)$ to all $x$ by a periodic continuation, whereas $\exp[i\pi x/(2\epsilon)]$ has a period of two cells. These observations allow us to rewrite \eqref{WKB asym} as
\begin{multline}\label{WKB asym reduced}
u \sim \frac{\exp{(ik_t x/\epsilon)}U_t(\xi)}{|x-x_t|^{1/4}\sqrt{\left|k_{1/2}\left(\pd{^2\Omega}{k^2}\right)_t\right|\int{\chi_t}|U_t|^2\,d\xi}} \times \\ \left\{\mathcal{A}_+ e^{i\gamma_t^+ + i\varphi_t^+/\epsilon-ik_tx_t/\epsilon}
 \exp\left[\frac{2}{3}i k_{1/2}\frac{(x-x_t)^{3/2}}{\epsilon}\right] + \right. \\ \left. 
\mathcal{A}_- e^{i\gamma_t^- + i\varphi_t^-/\epsilon+ik_tx_t/\epsilon-2i\beta}\exp\left[ - \frac{2}{3}i k_{1/2}\frac{(x-x_t)^{3/2}}{\epsilon}\right] \right\} \quad \text{as} \quad x\to x_t.
\end{multline}
We now see that the field $u$ is asymptotic to an oscillating tale of an Airy function, varying on an intermediate $O(\epsilon^{2/3})$ scale, and modulated on the short $O(\epsilon)$  scale by the periodic or anti-periodic standing-wave Bloch eigenfunction $\exp{(ik_t x/\epsilon)}U_t(\xi)$. On one hand, an Airy tale is reminiscent of the behaviour of a single-scale WKB solution near a turning point \cite{Holmes:Book}. On the other hand, a multiple-scale solution where an envelope function is modulated by a short-scale standing-wave Bloch eigenmode is nothing but the ansatz assumed in HFH \cite{Craster:10}. These two observations together suggest looking next at the region $x-x_t=O(\epsilon^{2/3})$ using a localised HFH-like expansion.

\section{High-frequency homogenisation of the band-gap edge}\label{sec:hfh}
Thus near the band-gap edge $x=x_t$ we look for a multiple-scale solution in the form
\begin{equation}\label{hfh ansatz}
u = \epsilon^{-1/6}e^{ik_t x/\epsilon}T(\xi,\bar{x}), \quad \bar{x}=\frac{x-x_t}{\epsilon^{2/3}},
\end{equation}
where $T$ is $2$-periodic in $\xi$, and $k_t=0$ or $\pi/2$. In terms of $\bar{x}$, 
\begin{equation}
\Lambda \sim \chi_t(\xi)+\epsilon^{2/3}\bar{x}\left(\frac{dr}{dx}\right)_t\left(\pd{\chi}{r}\right)_t(\xi) + O(\epsilon);
\end{equation}
note that $\tau$ enters only at $O(\epsilon)$ and will not affect the leading order of $T$. 
Substitution of \eqref{hfh ansatz} into the governing equation \eqref{master eq} gives
\begin{equation}\label{T eq}
\mathscr{L}^2_{k_t,\Lambda}T+ \epsilon^{1/3}\left(2\pd{^2T}{\bar{x}\partial\xi}+2ik_t\pd{T}{\bar{x}}\right) + \epsilon^{2/3}\pd{^2T}{\bar{x}^2} = 0.
\end{equation}
The form of \eqref{T eq} suggests the expansion
\begin{equation}
T(\xi,\bar{x}) \sim T_0(\xi,\bar{x}) + \epsilon^{1/3}T_1(\xi,\bar{x}) + \epsilon^{2/3}T_2(\xi,\bar{x})+ \cdots,
\end{equation}
where as usual $T_0,T_1,\ldots$ are 2-periodic in $\xi$. 
The $O(1)$ balance of \eqref{T eq},
\begin{equation}\label{T O1}
\mathscr{L}^2_{k_t,\chi_t}T_0=0,
\end{equation}
in conjunction with periodicity on the short scale, is identified as the zone-centre ($k_t=0$) or zone-edge ($k_t=\pi/2$) Bloch problem. The solution \eqref{T O1} therefore reads
\begin{equation}
T_0 = g_0(\bar{x})U_t(\xi),
\end{equation}
where $g_0$ is a leading-order envelope function, and the standing-wave Bloch eigenfucntion $U_t(\xi)=U[\xi;(k_t,r_t)]$ is specifically chosen as in \S\ref{sec:behaviour}.  The $O(\epsilon^{1/3})$ balance of \eqref{T eq} gives
\begin{equation}
\mathscr{L}^2_{k_t,\chi_t}T_1=-2\frac{dg_0}{dx}\left(ik_tU_t+\frac{dU_t}{d\xi}\right).
\end{equation}
It is readily verified that solvability is identically satisfied. The general solution is
\begin{equation}
T_1 = g_1(\bar{x})U_t(\xi)+\frac{dg_0}{d\bar{x}}\Phi(\xi),
\end{equation}
where $\Phi$ is any particular solution of 
\begin{equation}
\mathscr{L}^2_{k_t,\chi_t}\Phi=-2ik_tU_t-2\frac{dU_t}{d\xi}
\end{equation}
that is $2$-periodic. 
Finally, consider the $O(\epsilon^{2/3})$ balance of \eqref{T eq}, 
\begin{multline}
\mathscr{L}^2_{k_t,\chi_t}T_2=-2\frac{dg_1}{dx}\left(ik_tU_t+\frac{dU_t}{d\xi}\right) \\
-\frac{d^2g_0}{d\bar{x}^2}\left(2\frac{d\Phi}{d\xi}+2ik_t\Phi_1+U_t\right)-\omega^2\bar{x}g_0\left(\frac{dr}{dx}\right)_t\left(\pd{\chi}{r}\right)_t U_t.
\end{multline}
A solvability condition is derived in the usual way. We find that the envelope function $g_0(\bar{x})$ is governed by an Airy equation,
\begin{equation}\label{scary Airy}
\frac{d^2g_0}{d\bar{x}^2}+\left[\frac{\omega^2\left(\frac{dr}{dx}\right)_t\int \left(\pd{\chi}{r}\right)_t |U_t|^2\,d\xi}{2\int \frac{d\Phi}{d\xi}U_t^*\,d\xi+2ik_t\int\Phi U^*_t\,d\xi+\int|U_t|^2\,d\xi}\right]\,\bar{x}g_0 =0,
\end{equation}
where the constant in the square brackets is provided in terms of integrals of the leading- and first-order short-scale eigenfunctions $U_t$ and $\Phi$. Remarkably, however, this constant turns out to be equal to $k^2_{1/2}$ [cf.~\eqref{k12 thermo}]; this connection is derived in the appendix, see \eqref{k12}. The Airy equation \eqref{scary Airy} therefore reduces to 
\begin{equation}\label{Airy}
\frac{d^2g_0}{d\bar{x}^2}+k_{1/2}^2\bar{x}g_0 =0.
\end{equation}
Hence
\begin{equation}\label{g solution}
g_0(\bar{x})=\mathcal{A}\,\mathrm{Ai}(\pm |k_{1/2}|^{2/3}\bar{x})+\mathcal{B}\,\mathrm{Bi}(\pm |k_{1/2}|^{2/3}\bar{x}),
\end{equation}
where $\mathcal{A}$ and $\mathcal{B}$ are constants, $\mathrm{Ai}$ and $\mathrm{Bi}$ are the first and second Airy functions, and the plus (minus) sign applies if crossing into the band gap as $x$ is increased (decreased).

We shall see that the leading-order multiple-scale solution $\epsilon^{-1/6}e^{ik_t x/\epsilon}U_t(\xi)g_0(\bar{x})$, with $g_0$ given by \eqref{g solution}, can be matched, on the pass-band side of $x_t$, with the propagating WKB solutions \eqref{WKB solution}. On the stop-band side of $x_t$, where $k$ is complex, we anticipate that the solution can be matched with evanescent WKB solutions, though we do not consider this here. The evanescent WKB waves may be ignored when tunnelling effects are absent or negligible and in what follows we consider one such scenario in detail.

\begin{figure}[t]
\begin{center}
\includegraphics[scale=0.55]{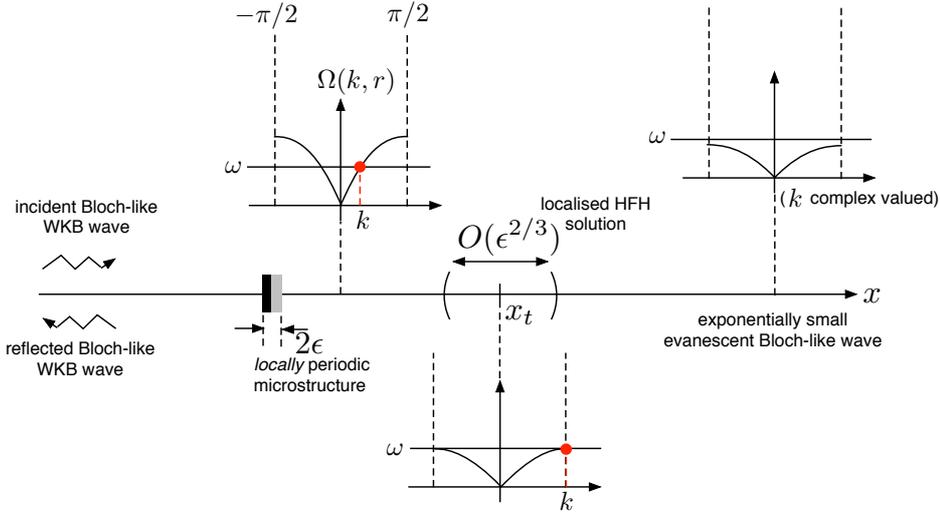}
\caption{Reflection from a spatial band-gap edge. An incident Bloch wave, described by a multiple-scale WKB approximation,  adiabatically propagates to the right. Adiabatic propagation breaks down as the Bloch wavenumber $k$ approaches $k_t=\pi/2$. Asymptotically matching the WKB approximation with a multiple-scale HFH solution localised about the singularity determines the reflected Bloch wave propagating left from the singularity.
}
\label{fig:reflection_schematic}
\end{center}
\end{figure}

\section{Bloch-wave reflection from a band-gap edge}\label{sec:reflection}
Consider a slowly varying Bloch wave incident on a band-gap edge singularity, assuming that the band gap is spatially wide so that tunnelling-assisted transmission across it is absent or of exponentially small order (see Fig.~\ref{fig:reflection_schematic}). We accordingly set $\mathcal{B}=0$ in \eqref{g solution}. The leading-order approximation in the transition region is therefore
\begin{equation}\label{trans leading}
u \sim \epsilon^{-1/6}\mathcal{A}e^{ik_t x/\epsilon}U_t(\xi) \mathrm{Ai}\left(\pm |k_{1/2}|^{2/3}\bar{x}\right),
\end{equation}
where the $\pm$ sign implies matching with the propagating WKB solution \eqref{WKB solution} in the limit as $\bar{x}\to\mp\infty$. We carry out the matching using van Dyke's rule \cite{Van:Book} with respect to the long-scale variables $x$ and $\bar{x}$, keeping the short-scale variable $\xi$ fixed. Thus, rewriting  \eqref{trans leading} in terms of $x$ rather than $\bar{x}$, then expanding to leading order in $\epsilon$, yields
\begin{multline}\label{trans expansion}
u \sim \mathcal{A}\frac{e^{ik_t x/\epsilon}U_t(\xi)}{2i\sqrt{\pi}|k_{1/2}|^{1/6}|x-x_t|^{1/4}}\left[e^{i\frac{\pi}{4}}e^{i\frac{2}{3}|k_{1/2}||x-x_t|^{3/2}/\epsilon} \right. \\ \left. - e^{-i\frac{\pi}{4}}e^{-i\frac{2}{3}|k_{1/2}||x-x_t|^{3/2}/\epsilon}\right].
\end{multline}
We compare \eqref{trans expansion} to the leading-order WKB solution \eqref{WKB solution}, rewritten in terms of $\bar{x}$ rather than $x$, then expanded to leading order in $\epsilon$. We thereby see that \eqref{trans expansion} and \eqref{WKB asym reduced} must coincide. 

Consider first the case where the band gap is to the right of $x_t$. If $k_t=0$ then $k>0$ for $x-x_t<0$ and hence $k_{1/2}=-i|k_{1/2}|$. Alternatively, if $k_t=\pi/2$, then $k-k_t<0$ for $x-x_t<0$ and hence $k_{1/2}=i|k_{1/2}|$; note also that $|x-x_t|^{3/2}=i(x-x_t)^{3/2}$. 
Thus matching yields the connection formulae 
\begin{gather}\label{connection1}
\frac{\mathcal{A}e^{i\frac{\pi}{4}}}{2i\sqrt{\pi}|k_{1/2}|^{1/6}} = \frac{\mathcal{A}_{\mp} e^{i\gamma_t^{\mp} + i\varphi_t^{\mp}/\epsilon\pm ik_tx_t/\epsilon-(1\pm 1)i\beta}}{\sqrt{\left|k_{1/2}\left(\pd{^2\Omega}{k^2}\right)_t\right|\int{\chi_t}|U_t|^2\,d\xi}}, \\ \label{connection2}
-\frac{\mathcal{A}e^{-i\frac{\pi}{4}}}{2i\sqrt{\pi}|k_{1/2}|^{1/6}}= \frac{\mathcal{A}_{\pm} e^{i\gamma_t^{\pm} + i\varphi_t^{\pm}/\epsilon\mp ik_tx_t/\epsilon-(1\mp1)i\beta}}{\sqrt{\left|k_{1/2}\left(\pd{^2\Omega}{k^2}\right)_t\right|\int{\chi_t}|U_t|^2\,d\xi}},
\end{gather}
where here the upper sign is for $k_t=0$ and the lower sign is for $k_t=\pi/2$. 
Similar considerations show that \eqref{connection1} holds also in the case where the band gap is to the left of $x_t$.
Either $\mathcal{A}_+$ or $\mathcal{A}_-$, whichever corresponds to the incident WKB wave propagating towards the singularity, is assumed known. The  other coefficient, determining the reflected WKB wave, is found by eliminating $\mathcal{A}$ from \eqref{connection1} and \eqref{connection2}:
\begin{equation}\label{reflection}
\frac{\mathcal{A}_+}{\mathcal{A}_- }= {-ie^{i(\gamma_t^- -\gamma_t^+)+i(\varphi_t^- -\varphi_t^+)/\epsilon+2ik_tx_t/\epsilon-2i\beta} }.
\end{equation}
The incident and reflected waves in \eqref{WKB solution} are  equal in amplitude up to the pre-factors $\mathcal{A}_+$ and $\mathcal{A}_-$. Thus \eqref{reflection} confirms that the incident wave is totally reflected and determines the initial phase of the reflected wave. Finally, obtaining $\mathcal{A}$ from either \eqref{connection1} or \eqref{connection2} determines the solution in the transition region.

\begin{figure}[t]
\begin{center}
\includegraphics[scale=0.3]{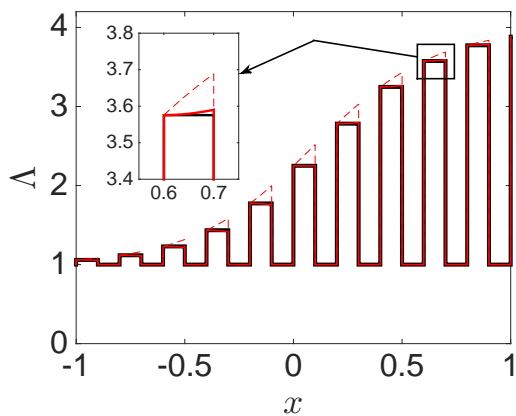}
\caption{Two-variable representation of the locally periodic layered medium considered in \S\ref{sec:layered}, in the case $r=1.5 +0.5\tanh(3x/2)$ and $\epsilon=0.1$. The black line depicts the material parameter $\Lambda$ exactly. The red dashed line shows the leading-order approximation $\Lambda\sim \chi(\xi,r)$, which is insufficiently accurate to serve the leading-order WKB approximation developed in this paper. The red solid line depicts the improved and sufficient approximation $\Lambda\sim\chi(\xi,r)+\epsilon\tau(\xi,r)$.}
\label{fig:layer_model}
\end{center}
\end{figure}
\section{Example: A layered medium}
\label{sec:layered}
\subsection{Consistent two-variable representation}
We now describe an implementation of the above asymptotic theory in the case of a locally periodic layered medium. Specifically, consider a layered medium made out of unit cells of width $2\epsilon$, where 
in the left half of each cell $\Lambda=1$ whereas in the right-half $\Lambda$ equals the value of $r^2(x)$ \emph{evaluated at the centre of that cell} (see black line in Fig.~\ref{fig:layer_model}). It is tempting to set $\Lambda = \chi[\xi,r(x)]$, where
\begin{equation}\label{layered chi}
\chi =\begin{cases}
1, & -1<\xi<0 \\
r^2,              & 0<\xi<1.
\end{cases}
\end{equation}
As prompted in \S\ref{sec:bulk}, however, this choice would not accurately represent our layered medium (see dashed red line in Fig.~\ref{fig:layer_model}). Indeed, $r(x)$ varies by $O(\epsilon)$ across the positive-$\xi$ half of each cell, while the true layered medium consists of strictly homogeneous layers. For this reason we allowed $\Lambda$ to have an $O(\epsilon)$ correction [cf.~\eqref{Lambda def}]. Specifically, a layered medium can be consistently modelled as $\Lambda = \chi[\xi,r(x)] + \epsilon \tau[\xi,r(x)] + O(\epsilon^2)$, where $\chi$ is provided by \eqref{layered chi} and
\begin{equation}\label{layered tau}
\tau =\begin{cases}
0, & -1<\xi<0 \\
-2\xi r\frac{dr}{dx},              & 0<\xi<1.
\end{cases}
\end{equation}
It is easy to see that the corrected material coincides, to $O(\epsilon)$, with the true layered medium (see red line in Fig.~\ref{fig:layer_model}). Our asymptotic theory reveals that the correction $\epsilon\tau$ is crucial in propagating the slow Berry-like phases $\gamma^{\pm}(x)$ and accordingly plays an important role in fixing the crests and troughs of the WKB Bloch waves; in contrast, the correction was not important in the leading-order HFH analysis of the transition region. 

\begin{figure}[t]
\begin{center}
\includegraphics[scale=0.37]{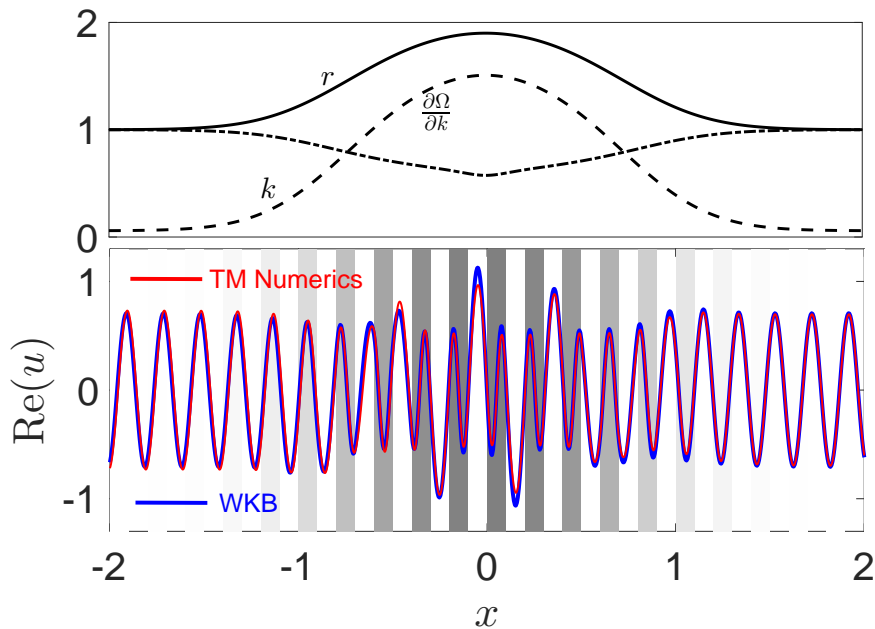}
\caption{Comparison between the WKB approximation and a TM simulation in the scenario where a Bloch-like wave propagates to the right through a locally periodic medium ($r=1.9-0.9\tanh(x^2),\, \epsilon=0.1,\,\omega=3.2$). The medium is homogeneous for $|x|\gg1$ and an incident plane wave is prescribed at large negative $x$. The weak reflection observed in the TM simulation is neglected in the adiabatic WKB approximation.} 
\label{fig:ad1}
\end{center}
\end{figure}
\begin{figure}[t]
\begin{center}
\includegraphics[scale=0.37]{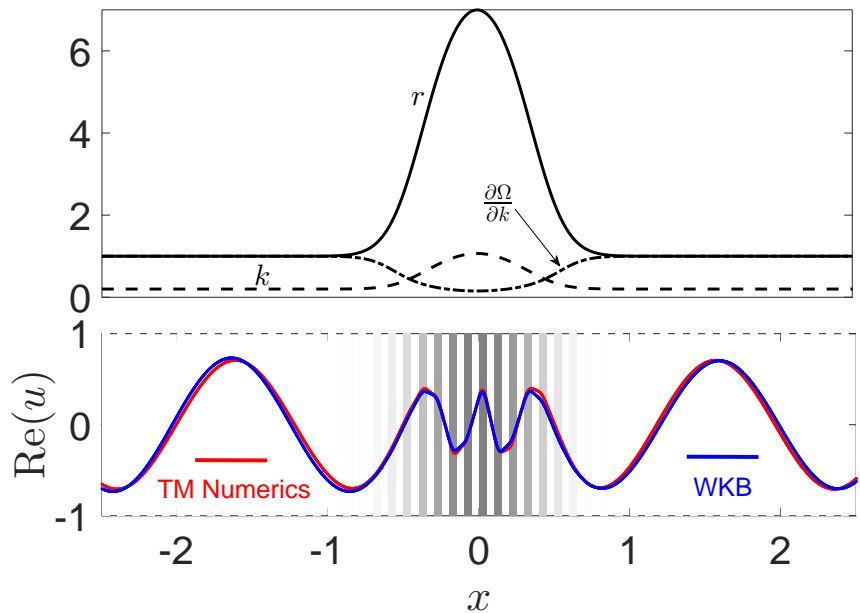}
\caption{
Like Fig.~\ref{fig:ad1} but for $r=7-6\tanh(4x^2),\,\epsilon=0.05,\,\omega=0.2$. } 
\label{fig:ad2}
\end{center}
\end{figure}
\begin{figure}[t]
\begin{center}
\includegraphics[scale=0.37]{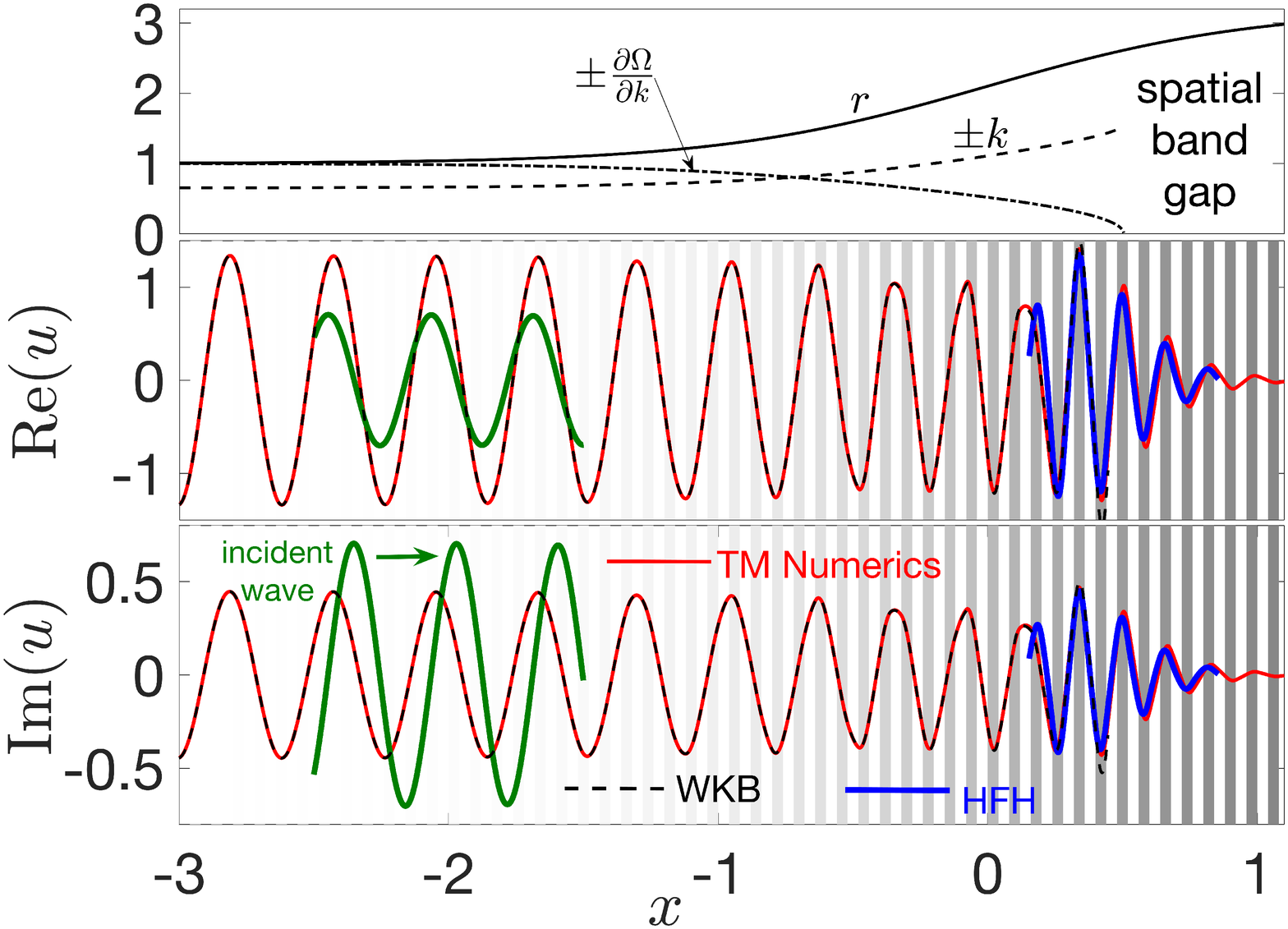}
\caption{
A Bloch-like wave, propagating to the right through a locally periodic medium ($r=2.1+1.1\tanh x,\, \epsilon=0.04,\,\omega=0.65$), is totally reflected from a spatial band gap. An incident plane wave is prescribed at large negative $x$ where the medium is homogeneous. A comparison is shown between a TM simulation and an asymptotic approximation compound of WKB waves matched to a HFH solution localised about the band-gap edge.  }
\label{fig:nonad}
\end{center}
\end{figure}

\subsection{Bloch eigenvalue problem}
To implement the asymptotic scheme we need to solve the Bloch eigenvalue problem at fixed $\omega$, as a function of the material property $r$ varying on the long scale. That problem consists of the differential equation $\mathscr{L}^2_{k,\chi}U=0$ together with periodic boundary conditions on the cell boundaries $\xi=\pm1$. The solution when $\chi$ is given by \eqref{layered chi} is well known  \cite{Kittel:Book}. In particular, the dispersion relation is 
\begin{equation}\label{layer dispersion}
2r\cos(2k)=2r\cos\omega\cos(r\omega)-(1+r^2)\sin\omega\sin(r\omega),
\end{equation}
while the periodic Bloch eigenfunctions can be written as
\begin{equation}\label{mode}
U(\xi,\{r,k\})=e^{-ik\xi}
\begin{cases}
a_1e^{i\omega\xi}+a_2e^{-i\omega\xi}, & -1<\xi<0 \\
  b_1e^{i\omega r\xi}+b_2e^{-i\omega r\xi},              & 0<\xi<1.
\end{cases}
\end{equation}
We fix the gauge by setting $a_1=e^{i\omega}\left[1+2e^{ir\omega}(r-1)+r\left(1-2e^{i(2k+\omega+\omega r)}\right)\right]$; 
the remaining coefficients $a_2$, $b_1$ and $b_2$ are then determined from periodicity along with continuity of $U$ and its first derivative at $\xi=0$. 

\subsection{Numerical validation: Adiabatic propagation}
In what follows we compare our asymptotic approach with a numerical transfer-matrix (TM) scheme \cite{Yeh:Book}.  
This numerical scheme takes advantage of the fact that within each layer the solution of \eqref{master eq} is a superposition of two harmonic waves propagating in opposite directions. The solution can therefore be represented by two layer-dependent complex coefficients that characterise the amplitudes and phases of the harmonic waves; if the two coefficients are known for one layer, the solution everywhere can be determined by applying continuity conditions at successive interfaces. Usually one does not know \emph{a priori} the values of both coefficients in any given cell and a shooting method is used.

We first demonstrate propagation in a locally periodic material, which, at a given operating frequency, is free of dispersion singularities. Accordingly, we expect the WKB solutions \eqref{WKB solution} to hold for all $x$. For the sake of simplicity we assume $r\to1$ as $x\to\pm\infty$, which means that at infinity the medium is homogeneous and accordingly the solution tends to a superposition of oppositely propagating plane waves. To simulate an incident wave from the left, we numerically prescribe at a large negative $x$ the coefficient of the right-propagating wave and determine the coefficient of the left-propagating wave by a shooting method that ensures the radiation condition is satisfied at large positive $x$ (no incoming plane wave). The corresponding asymptotic solution is given by  the right-propagating  WKB Bloch wave in \eqref{WKB solution}, which is advanced using \eqref{phi pm} and \eqref{gamma pm} starting from the incident plane wave at large negative $x$; the asymptotic scheme does not predict a reflected Bloch wave in this case hence the second solution in \eqref{WKB solution} is discarded.  

Figs.~\ref{fig:ad1} and \ref{fig:ad2} show a comparison between the WKB asymptotics and the TM scheme in the respective cases (i) $\omega=3.2,\,\epsilon=0.1,\,r=1.9-0.9\tanh(x^2)$ and (ii) $\omega=0.2,\,\epsilon=0.05,\,r=7-6\tanh(4x^2)$. The figures also show the variation of the material parameter $r$ along with the local Bloch wavenumber and group velocity. The agreement between the asymptotic and numerical solutions is excellent. Fig.~\ref{fig:ad1} shows that the method remains accurate even when the long scale is not very large compared with the short scale. Moreover, at $x=0$ the local Bloch eigenvalue is close to a branch crossing and this too does not result in appreciable deviations. The comparison in Fig.~\ref{fig:ad2} demonstrates the efficacy of the WKB solution for strong and moderately rapid variations. The frequency $\omega$ is low, which explains the long wavelength for $|x|\gtrsim1$. For $|x|\lesssim 1$, however, $r$ goes up to $7$ and the wavelength is intermittently short and long in the right and left halves of each unit cell (the wave field is accordingly curved in the grey layers and almost linear in the white layers). Note also that at $x=0$ the Bloch eigenvalue is close to a band-gap edge, where the WKB description breaks down. Thus we have demonstrated agreement under extreme circumstances. We stress that if a naive approach is taken, where  either the slow phase $\gamma$ or the medium correction $\tau$ are ignored, the agreement is quite poor and the error does not vanish as $\epsilon\to0$. 

\subsection{Numerical validation: Reflection from a band-gap edge}
We next demonstrate the scenario discussed in \S\ref{sec:reflection} where a Bloch wave is reflected from a band-gap edge. Specifically, we consider a locally periodic layered medium $r=2.1+1.1\tanh x$. The medium is homogeneous at large negative $x$, slowly transforming into a band-gap medium at larger $x$; for an operating frequency $\omega=0.65$ the Bloch wavenumber $k$ is real for $x<x_t\approx 0.5$ and complex  for $x>x_t$ with real part $k_t=\pi/2$ (see  Fig.~\ref{fig:nonad}). To construct the asymptotic solution we advance the right-propagating WKB wave from a prescribed incident wave at large negative $x$. We then use the theory in \S\ref{sec:behaviour}--\S\ref{sec:reflection} to match the diverging solution with a local HFH description for $|x-x_t|=O(\epsilon^{2/3})$, which in turn determines through matching initial conditions for advancing a reflected left-propagating WKB wave. The numerical scheme is applied as before but now with a shooting method ensuring there is no exponentially growing evanescent wave for $x>x_t$. Fig.~\ref{fig:nonad} shows excellent agreement between the asymptotic and numerical solutions for $\epsilon=0.04$. We note that  $\gamma$ and $\tau$ play a particularly crucial role in this scenario since matching sensitively depends on the limiting value $\gamma_t$ of the slow phase. We lastly note that in principle the HFH solution could also be matched with an exponentially decaying  evanescent Bloch wave for $x>x_t$, though this will not affect the reflected wave in the present scenario where tunnelling is absent.

\section{Discussion}\label{sec:discussion}
The key message of this paper is that multiple-scale WKB approximations and HFH are complementary and have overlapping domains of validity. The two methods can therefore be systematically combined, through the method of matched asymptotic expansions, to furnish a more complete scheme for high-frequency waves in periodic, or locally periodic, media. We here demonstrated this idea for a locally periodic medium in 1D, focusing on the scenario of total reflection of a slowly varying Bloch wave from a band-gap edge. Clearly this is only a first step towards a more general theory.

The present 1D analysis can be directly applied to calculate trapped modes for a locally periodic medium with a pass-band interval bounded by two stop-band intervals \cite{Johnson:02}. By generalising the analysis to account for off-axis propagation, the latter calculation could be adapted to model Bragg waveguides \cite{Yeh:76}. An analogous 1D analysis could be devised to describe transmission and reflection of WKB Bloch waves in the presence of degenerate points where two dispersion branches touch; this would entail matching the WKB solutions with modified inner regions treated by a degenerate HFH-like expansion \cite{Colquitt:15high}. It would also be useful to generalise the WKB scheme to include evanescent Bloch waves; one could then describe tunnelling-assisted transmission through a locally periodic medium with a stop-band interval bounded by two pass-band intervals. Further objectives, still in the 1D framework, include implementing the method for additional periodic modulations $\Lambda$ and to generalise the asymptotics to other wave equations; in particular, in electro-magnetics the wave equation \eqref{master eq} describes s-polarised light and it would be desirable to study p-polarised light as well. 

In higher dimensions the multiple-scale WKB method generalises to a geometric ray theory \cite{Allaire:11}.  We anticipate that WKB solutions could still be matched with HFH-like expansions localised about geometric places where the Bloch eigenvalue becomes critical or degenerate. Spatially localised dispersion singularities would be present even in non-graded crystals, since the Bloch wavevector at fixed frequency would vary with the direction of propagation. In the latter case, the Bloch wavevector is conserved along straight rays that are directed along the corresponding group-velocity vector. HFH expansions would thus be sought about singular rays whose Bloch wavevector is critical. In higher dimensions there would also be the usual ray singularities, in physical rather than Bloch space, namely focus points and caustics \cite{Chapman:99}. Finally, a general description would address diffraction at crystalline interfaces. While Bragg's law determines the directions of reflected, refracted and diffracted rays, a complete quantitative picture would generally entail matching with inner regions wherein the crystal appears to be semi-infinite. 

\section*{Acknowledgement}
I am grateful to Richard V. Craster for fruitful discussions. 

\appendix
\section{Perturbation theory applied to the Bloch eigenvalue problem}\label{app:freq_pert}
In \S\ref{sec:bulk} we formulated the Bloch problem for the wavenumber $k(x)$ and eigenfunction $U(\xi,\{k(x),r(x)\})$ at fixed frequency $\omega$ and position $x$. Here we analyse small perturbations of this problem as a means to derive \emph{exact} identities satisfied by the Bloch eigenvalues and eigenfunctions. 
\subsection{Perturbation with $x$ fixed}
Say $x$ is fixed and consider small perturbations of the eigenfrequency $\Omega$ from $\omega$ [cf.~\eqref{disp}].  We denote the perturbed wavenumber and eigenfunction by $\acute{k}$ and $\acute{U}$ such that $\acute{k}=k$ and $\acute{U}=U[\xi,\{k,r\}]$ when $\Omega=\omega$. The perturbed eigenvalue problem consists of the differential equation
\begin{equation}\label{app bloch}
\pd{^2\acute{U}}{\xi^2}+2i\acute{k}\pd{\acute{U}}{\xi}+\left(\Omega^2{\chi}-\acute{k}^2\right)\acute{U}=0,
\end{equation}
in conjunction with the condition that $\acute{U}$ satisfies periodic conditions at $\xi=\pm1$. Introducing the small parameter $\delta=\acute{k}-k$, we expand the eigenfrequency and eigenfunction as
\begin{equation}\label{expansions}
\Omega \sim \omega + \delta \pd{\Omega}{k}+O(\delta^2), \quad\acute{U}\sim U + \delta \acute{U}_1 + O(\delta^2).
\end{equation}
We now substitute \eqref{expansions} into \eqref{app bloch} and compare respective orders in $\delta$. By construction, at $O(1)$ we simply find the Bloch problem satisfied by $U$. At $O(\delta)$ we find 
\begin{equation}
\pd{^2\acute{U}_1}{\xi^2}+2ik\pd{\acute{U}_1}{\xi}+\left(\omega^2\chi-k^2\right)\acute{U}_1 = -2i\pd{U}{\xi}-\left(2\omega\pd{\Omega}{k}\chi-2k\right)U.
\end{equation}
Deriving a solvability condition using the adjoint problem governing $U^*$ gives 
\begin{equation}\label{gv relation app}
\pd{\Omega}{k}= \frac{k\int|U|^2\,d\xi-i\int  U^*\pd{U}{\xi}\,d\xi}{\omega\int\chi|U|^2\,d\xi}.
\end{equation}
It is indeed well known that the group velocity can be expressed in terms of integrals of the eigenfunction \cite{Sakoda:Book}, which circumvents the need to differentiate the dispersion relation \eqref{disp}. 

\subsection{Perturbation with $\omega$ fixed}
\label{app:x_pert}
\subsubsection{Pass band} \label{ssec:pass}
We next fix $\omega$ and consider small perturbations of $x$ from a given value $x'$. Since $\omega$ is fixed the situation conforms to the notation used in the main text, where the Bloch problem consists of $\mathscr{L}^2_{k,\chi}U=0$ [cf.~\eqref{L2}] together with periodicity conditions at $\xi=\pm1$. We first consider the case where $0<k(x')<\pi/2$ is real and non-degenerate, in which case we anticipate the expansions
\begin{gather}
k \sim k'+ \left(\frac{dk}{dx}\right)'(x-x') + O[(x-x')^2], \\
\chi \sim \chi'+ \left(\frac{dr}{dx}\right)'\left(\pd{\chi}{r}\right)'(x-x') + O[(x-x')^2],\\
U \sim U'+\left(\pd{U}{x}\right)'(x-x')+ O[(x-x')^2],
\end{gather}
where a prime denotes evaluation at $x=x'$. 
By construction, the leading-order problem is identically satisfied. At order $O(x-x')$ we find
\begin{equation}\label{pert1}
\mathscr{L}^2_{k',\chi'}\left(\pd{U}{x}\right)'=-2i\left(\frac{dk}{dx}\right)'\pd{U'}{\xi}-\omega^2\left(\frac{dr}{dx}\right)'\left(\pd{\chi}{r}\right)'U'+2k'\left(\frac{dk}{dx}\right)'U'.
\end{equation}
The solvability condition on \eqref{pert1}, when separated to its real and imaginary parts, yields the following exact relations (dropping the primes):
\begin{gather}\label{dk}
\frac{dk}{dx}=\frac{\omega}{2}\frac{dr}{dx}\left(\pd{\Omega}{k}\right)^{-1}\frac{\int\pd{\chi}{r}|U|^2\,d\xi}{\int\chi|U|^2\,d\xi},\\
\label{dU}
\mathrm{Re}\int \left(\pd{U}{\xi}+ikU\right)\pd{U^*}{x}\,d\xi=-\frac{\omega^2}{2}\left(\frac{dk}{dx}\right)^{-1}\frac{dr}{dx}\mathrm{Im}\int\pd{\chi}{r}\left(\pd{U^*}{x}U\right)\,d\xi.
\end{gather}

\subsubsection{Perturbation from a Band-gap edge} \label{ssec:BG_pert}
We next consider small perturbations of $x$ from a band-gap edge $x_t$, where $r=r_t$, $k=k_t=0$ or $\pi/2$, and $U=U_t(\xi)$ (see \S\ref{sec:behaviour}). The appropriate expansions for $r$, $\chi$ and $k$ are now given by \eqref{r exp}, \eqref{chi exp} and \eqref{k exp}, respectively. Upon substituting these into $\mathscr{L}^2_{k,\chi}U=0$, it becomes clear that 
generally the expansion for $U$ possesses the form
\begin{equation}\label{caveat}
U\sim U_t +h(x-x_t)U_t + O[(x-x_t)^{1/2}], 
\end{equation}
where $h=o(1)$ but $\gg (x-x_t)^{1/2}$ as $x-x_t\to0$. The function $h$ depends on the gauge choice and we may assume it vanishes; this would be the case in any analytical or computational scheme where the eigenfunction is chosen to vary smoothly with $k$ [cf.~\eqref{k exp}]. We therefore write
\begin{equation}
U\sim U_t(\xi)+(x-x_t)^{1/2}U_{1/2}(\xi) + (x-x_t)U_1(\xi) + \cdots.
\end{equation}
At $O(x-x_t)^{1/2}$ we find
\begin{equation}\label{be 12}
\pd{^2U_{1/2}}{\xi^2}+2i k_t\pd{U_{1/2}}{\xi} + (\omega^2\chi_t-k_t^2)U_{1/2}=-2ik_{1/2}\pd{U_t}{\xi}+2k_tk_{1/2}U_t.
\end{equation}
This forced equation automatically satisfies the solubility condition, since the inner product of the right hand side with $U_t^*$ is proportional to the vanishing group velocity at $x_t$ [cf.~\eqref{gv relation}]. The general solution to \eqref{be 12} can therefore be written as 
\begin{equation}
U_{1/2}=aU_t(\xi)+ik_{1/2}\Phi(\xi),
\end{equation}
where $\Phi$ is {any} periodic function satisfying
\begin{equation}\label{Phi eq}
\pd{^2\Phi}{\xi^2}+2i k_t\pd{\Phi}{\xi} + (\omega^2\chi_t-k_t^2)\Phi=-2\pd{U_t}{\xi}-2ik_tU_t,
\end{equation}
where the constant $a$ and the magnitude of the homogeneous solution of \eqref{Phi eq} depends on the gauge choice. 
We note that $\Phi\exp(ik_t\xi)$ is real up to a constant; in fact, since $\exp(i\beta)U_t\exp(ik_t\xi)$ is real than so is $\exp(i\beta)\Phi\exp(ik_t\xi)$. 
Finally, a solvability condition at $O(x-x_t)$ yields
\begin{equation}\label{k12}
k_{1/2}^2 = \frac{\omega^2\left(\frac{dr}{dx}\right)_t\int \left(\pd{\chi}{r}\right)_t |U_t|^2\,d\xi}{2\int \frac{d\Phi}{d\xi}U_t^*\,d\xi+2ik_t\int\Phi U^*_t\,d\xi+\int|U_t|^2\,d\xi},
\end{equation}
which is gauge invariant. 

\bibliographystyle{unsrt}
\bibliography{refs.bib}

\end{document}